\documentclass[conference,a4paper]{IEEEtran}
\IEEEoverridecommandlockouts

\usepackage{fancyhdr}
\usepackage[dvips]{graphicx}
\usepackage{indent}
\usepackage{amsmath,amssymb}
\usepackage{paralist}
\usepackage{eclbkbox}
\usepackage{subfigure}
\usepackage{multirow}

\begin{document}
\title{Emotion Orientated Recommendation System\\for Hiroshima Tourist by Fuzzy Petri Net
\thanks{\copyright 2013 IEEE. Personal use of this material is permitted. Permission from IEEE must be obtained for all other uses, in any current or future media, including reprinting/republishing this material for advertising or promotional purposes, creating new collective works, for resale or redistribution to servers or lists, or reuse of any copyrighted component of this work in other works.}
}

\author{\IEEEauthorblockN{Takumi Ichimura}
\IEEEauthorblockA{Faculty of Management and Information Systems,\\
Prefectural University of Hiroshima\\
1-1-71, Ujina-Higashi, Minami-ku,\\
Hiroshima, 734-8559, Japan\\
Email: ichimura@pu-hiroshima.ac.jp}
\and
\IEEEauthorblockN{Issei Tachibana}
\IEEEauthorblockA{Graduate School of Comprehensive Scientific Research,\\
Prefectural University of Hiroshima\\
1-1-71, Ujina-Higashi, Minami-ku,\\
Hiroshima, 734-8559, Japan\\
Email: isseing1224@gmail.com}
}

\maketitle

\fancypagestyle{plain}{
\fancyhf{}	
\fancyfoot[L]{}
\fancyfoot[C]{}
\fancyfoot[R]{}
\renewcommand{\headrulewidth}{0pt}
\renewcommand{\footrulewidth}{0pt}
}

\pagestyle{fancy}{
\fancyhf{}
\fancyfoot[R]{}}
\renewcommand{\headrulewidth}{0pt}
\renewcommand{\footrulewidth}{0pt}

\begin{abstract}
We developed an Android Smartophone application software for tourist information system. Especially, the agent system recommends the sightseeing spot and local hospitality corresponding to the current feelings. The system such as concierge can estimate user's emotion and mood by Emotion Generating Calculations and Mental State Transition Network. In this paper, the system decides the next candidates for spots and foods by the reasoning of fuzzy Petri Net in order to make more smooth communication between human and smartphone. The system was developed for Hiroshima Tourist Information and described some hospitality about the concierge system.
\end{abstract}


\IEEEpeerreviewmaketitle

\section{Introduction}
\label{sec:Introduction}
Our research group proposed an estimation method to calculate the agent's emotion from the contents of utterances and to express emotions which are aroused in computer agent by using synthesized facial expression \cite{Ichimura03, Mera02, Mera03}. Emotion Generating Calculations (EGC) method \cite{Mera03} based on the Emotion Eliciting Condition Theory \cite{Elliott92} can decide whether an event arouses pleasure or not and quantify the degree of pleasure under the event.

Calculated emotions by EGC will change the mood of the agent. Ren \cite{Ren06} describes Mental State Transition Network (MSTN) which is the basic concept of approximating to human psychological and mental responses. The assumption of discrete emotion state is that human emotion is classified into some kinds of stable discrete states, called ``mental state,'' and the variance of emotions occurs in the transition from a state to other state with an arbitrary probability. Mera and Ichimura \cite{Mera10, Ichimura13} developed a computer agent that can transit a mental state in MSTN based on analysis of emotion by EGC method. EGC calculates emotion and the type of the aroused emotion is used to transit mental state \cite{Mera10}.

 We have developed Android EGC application software which the agent works to evaluate the feelings in the conversation\cite{Ichimura12}. The smartphone user can not only obtain the variety of information but also converse with the agent in a smartphone, because the interface between human and smartphone has been equipped with the speech recognition. Moreover, smartphones have the GPS device and acceleration sensor. Our proposed technique, EGC, can be expected to be an emotional orientated interface. Moreover, our developed application is tourist information system which can estimate the user's feelings at the sightseeing spot. The system can recommend the sightseeing spot and the local food corresponded to the user's feeling. For example, the system works to guide the spot where the user feels ``happy'', even if he/she feels ``disgust.'' Our developed system for Hiroshima Tourist can guide some spots, local food shops, and local gifts described in Hiroshima Tourist Map Android application\cite{Android_Market}. In this paper, the system decides the next candidates for spots and foods by the reasoning of fuzzy Petri Net \cite{Chen91} in order to make more smooth communication between human and smartphone. We developed the fuzzy production rules to serve the hospitality in the concierge system. The rules react to the specified word in the user's utterances according to the emotion by using EGC and the mood at each spot.

The remainder of this paper is organized as follows. In the section \ref{sec:EGC}, the brief explanation to understand the EGC is described. Section \ref{sec:FuzzyPetriNet} describe the fuzzy Petri Net model. In Section \ref{sec:HiroshimaTouristInformation}, the recommendation system such concierge system is explained. In Section \ref{sec:ConclusiveDiscussion}, we give some discussions to conclude this paper.

\section{Emotion Generating Calculations}
\label{sec:EGC}
\subsection{An Overview of Emotion Generating Process}

\begin{figure}[ht]
\begin{center}
\includegraphics[scale=0.5]{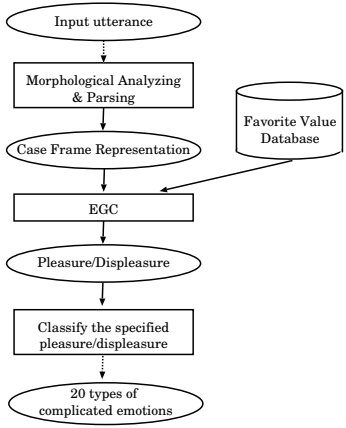}
\caption{Process for generating emotions}
\label{fig:processgeneratingemotion}
\end{center}
\end{figure}

Fig.\ref{fig:processgeneratingemotion} shows the emotion generating process where the user's utterance is transcribed into a case frame representation based on the results of morphological analysis and parsing. The agent works to determine the degree of pleasure/displeasure from the event in case frame representation by using EGC. EGC consists of 2 or 3 terms such as  subject, object and predicate, which have Favorite Value ($FV$), the strength of the feelings described in section \ref{sec:FavoriteValue}.

Then, the agent divides this simple emotion (pleasure/displeasure) into 20 various emotions based on the Elliott's ``Emotion Eliciting Condition Theory\cite{Elliott92}.'' Elliott's theory requires judging conditions such as ``feeling for another,'' ``prospect and confirmation,'' and ``approval/disapproval.'' The detail of this classification method is described in the section \ref{sec:ComplicatedEmotion}.

\subsection{Case Frame Representation}
\begin{figure}[!ht]
\begin{center}
\includegraphics[scale=0.8]{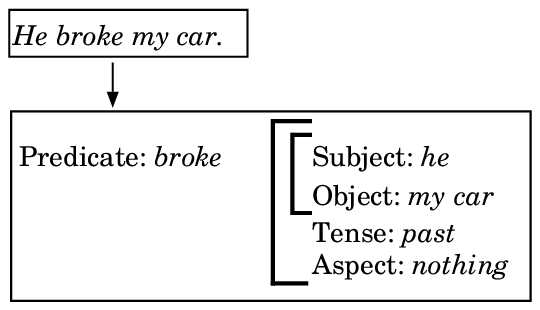}
\caption{Example of Case Frame Structure}
\label{fig:caseframerepresentation}
\end{center}
\end{figure}

The case frame structure bases the predicate phrase and the syntactic dependency between it and the other case elements. Fillmore\cite{Fillmore68} developed a system of linguistic analysis where the theory analyzes the surface syntactic structure of sentences by the combination of deep cases, e.g. semantic roles. Each verb selects a certain number of deep cases which form its case frame as shown in Fig.\ref{fig:caseframerepresentation}. 

In order to transcribe the user's utterances into the case frame representation, we implement morphological analysis and parsing to the input sentence.

\subsection{Favorite Value Database}
\label{sec:FavoriteValue}
Which an event is pleasure or displeasure is determined by using $FV$. $FV$ is a positive/negative number to an object when the user likes/dislikes it, respectively. $FV$ is predefined a real number in the range $[-1.0, 1.0]$. There are two types of $FV$s, personal $FV$ and initial $FV$. Personal $FV$ is stored in a personal database for each person who the agent knows well, and it shows the degree of like/dislike to an object from the person's viewpoint. On the other hand, an initial $FV$ shows the common degree of like/dislike to an object that the agent feels. Generally, it is generated based on the agent's own preference information according to the result of some questionnaires. Both personal and initial $FV$s are stored in the user own database. An initial value of $FV$ is determined beforehand on the basis of `corpus' of its applied field. The $FV$s of the objects are gained from a questionnaire. However, there are countless objects in the world. In this paper, we limit the objects that have initial $FV$ into the frequently appeared words in the dialog during sightseeing.

\subsection{Equation of EGC}
We assume an emotional space as three-dimensional space. Therefore, we present a method to distinguish pleasure/displeasure from an event by judging the existence of `synthetic vector''\cite{Mera02}.

\begin{table}[tbp]
\begin{center}
\caption{Correspondence between the event type and the axis}
\begin{tabular}{c|c|c|c}
\hline
Event type  & $f_{1}$ & $f_{2}$ & $f_{3}$ \\ \hline 
$V(S)$      &         &        &         \\
$A(S,C)$    &         &        &         \\
$A(S,OF,C)$ &         &        &         \\
$A(S,OT,C)$ & $f_{S}$  &        & $f_{P}$ \\
$A(S,OM,C)$ &         &        &         \\
$A(S,OS,C)$ &         &        &         \\ \hline
$V(S,OF)$   & $f_{S}$ & $f_{OT}-f_{OF}$ & $f_{P}$  \\
$V(S,OT)$   &         &         &         \\ \hline
$V(S,OM)$   & $f_{S}$ & $f_{OM}$ & $f_{P}$  \\ \hline
$V(S,OS)$   & $f_{S}-f_{OS}$ &  & $f_{P}$  \\ \hline
$V(S,O)$    & $f_{S}$ & $f_{O}$ & $f_{P}$  \\ 
            & $f_{O}$ &         & $f_{P}$ \\ \hline
$V(S,O,OF)$ & $f_{O}$ & $f_{OT}-f_{OF}$ & $f_{P}$ \\
$V(S,O,OT)$ &         & $f_{OM}$       &  \\ \hline
$V(S,O,OM)$ & $f_{O}$ & $f_{OM}$ & $f_{P}$ \\ \hline
$V(S,O,I)$ & $f_{O}$ & $\mid f_{I} \mid$  & $f_{P}$ \\ \hline
$V(S,O,OC)$ & $f_{O}$ &          & $f_{OC}$ \\ \hline
$A(S,O,C)$  & $f_{O}$ &          & $f_{P}$ \\ \hline
\end{tabular}
\label{tab:EGC-eventtype}
\end{center}
\end{table}

Table \ref{tab:EGC-eventtype} shows the correspondence between the case element in EGC equations and the axis in the three-dimensional model. In Table \ref{tab:EGC-eventtype}, `V(S,*)' is the type of event (verb) and `A(S,*)' is the type of attribute (adjective). the variables denoted in Table \ref{tab:EGC-eventtype} are expressed as follows.

\begin{itemize}
\item $f_{S}$ : $FV$ of Subject
\item $f_{OF}$ : $FV$ of Object-From
\item $f_{OM}$ : $FV$ of Object-Mutual
\item $f_{OC}$ : $FV$ of Object-Content
\item $f_{O}$ : $FV$ of Object
\item $f_{OT}$ : $FV$ of Object-To
\item $f_{OS}$ : $FV$ of Object-Source
\item $f_{P}$ : $FV$ of Predicate
\item $f_{I}$ : $FV$ of Instrument or tool
\end{itemize}

\begin{figure}[btp]
\begin{center}
\includegraphics[scale=0.35]{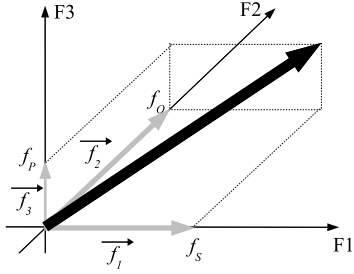}
\caption{Emotion Space for EGC}
\vspace{-0.5cm}
\label{fig:EGC-emotionvector}
\end{center}
\end{figure}

Table \ref{tab:EGC-axis} shows the relation between the sign of axis in each dimension and the pleasure/displeasure of generated emotion. When the vector is on the axis, the event does not raise any emotion. When we calculate the synthetic vectors of the events which do not have $f_{i}$ elements, we supply a dummy $FV$, $\beta$ as $f_{i}$ element. We tentatively defined $\beta$ as $+0.5$. Fig.\ref{fig:EGC-emotionvector} is an example of emotion space of event type $V(S, O)$. There are three elements, Subject, Object, and Predicate, in the event type, and the orthogonal vectors by the elements construct a rectangular solid. 

\begin{table}[tbp]
\begin{center}
\caption{pleasure/displeasure in emotional space}
\begin{tabular}{c|c|c|c|c}
\hline
Area & $f_{1}$ & $f_{2}$ & $f_{3}$ & Emotion \\ \hline
I & + & + & + & Pleasure \\
I\hspace{-.1em}I & - & + & + & Displeasure \\
I\hspace{-.1em}I\hspace{-.1em}I & - & - & + & Pleasure \\
I\hspace{-.1em}V & + & - & + & Displeasure \\
V & + & + & - & Displeasure \\
V\hspace{-.1em}I & - & + & - & Pleasure \\
V\hspace{-.1em}I\hspace{-.1em}I & - & - & - & Displeasure \\
V\hspace{-.1em}I\hspace{-.1em}I\hspace{-.1em}I & + & - & - & Pleasure \\ \hline
\end{tabular}
\label{tab:EGC-axis}
\end{center}
\end{table}

\subsection{Complicated Emotion Eliciting Method}
\label{sec:ComplicatedEmotion}
\begin{figure}[btp]
\begin{center}
\includegraphics[scale=0.7]{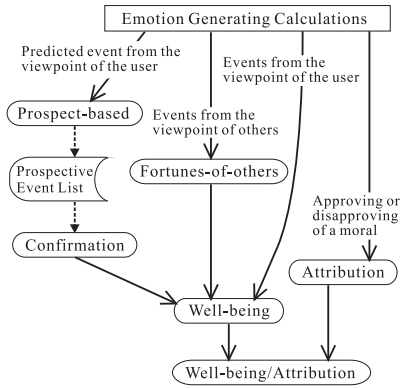}
\caption{Dependency among emotion groups}
\label{fig:EGC}
\end{center}
\end{figure}

Based on emotion values calculated by EGC method and their situations, the pleasure/displeasure is classified into 20 types of emotion. We consider only 20 emotion types, which are classified into six emotional groups as follows, ``joy'' and ``distress'' as a group of ``Well-Being,'' ``happy-for,'' ``gloating,'' ``resentment,'' and ``sorry-for'' as a group of ``Fortunes-of-Others,'' ``hope'' and ``fear'' as a group of ``Prospect-based,'' ``satisfaction,'' ``relief,'' ``fears-confirmed,'' and ``disappointment'' as a group of ``Confirmation,'' ``pride,'' ``admiration,'' ``shame,'' and ``disliking'' as a group of ``Attribution,'' ``gratitude,'' ``anger,'' ``gratification,'' and ``remorse'' as a group of ``Well-Being/Attribution'' \cite{Mera03}. Fig.\ref{fig:EGC} shows the dependency among the groups of emotion types. 

\section{Fuzzy Petri Net}
\label{sec:FuzzyPetriNet}
The interactive system for tourist concierge is implemented by the goal driven reasoning. The inference technique which uses IF-Then rules to repetitively breaks a goal into smaller sub-goals. It is an efficient way to solve problems that can be modeled as ``structured selection problems.'' The aim of the system is to pick the choice from many enumerated possibilities. 

The goal driven reasoning can deduce the user's requirement from the conversation in the specified reasoning system such as tourist information system. Then, IF-THEN rules for the interactive system as knowledge representations are prepared according to the sightseeing spots and goods.

A fuzzy production rule is a rule which describes the fuzzy relation between 2 proposition. Let $R$ be a set of fuzzy production rule $R=\{R_{1}, R_{2}, \cdots, R_{n}\}$. The general form of the $i$th fuzzy production rule $R_{i}$ is as follows:
\begin{equation}
R_{i}:IF\; d_{j}\; Then\; d_{k},\; CF=\mu_{j}
\label{eq:Rule-0}
\end{equation}

 A Fuzzy Petri Net (FPN)\cite{Chen91} is an effective model to implement goal driven reasoning. A FPN structure is defined as 8-tuple;
\begin{equation}
FPN = \{P, T, D, I, O, f, \alpha, \beta\},
\label{eq:FPN0-1}
\end{equation}
where $P=\{p_{1}, p_{2}, \cdots, p_{n}\}$ is a finite set of places, $T=\{t_{1}, t_{2}, \cdots, t_{m} \}$ is a finite set of transitions, $D=\{d_{1}, d_{2}, \cdots, d_{n} \}$ is a finite set of propositions, $P \cap T \cap D = \phi$ and $|P|=|D|$.  $I:T\rightarrow P^{\infty}$ is the input function, a mapping from transition to bags of places. $O: T\rightarrow P^{\infty}$ is the output function, a mapping from transition to bags of places. $f: T\rightarrow [0,1]$ is an association function, a mapping from transitions to real values in [0, 1]. $\alpha: P\rightarrow [0,1]$ is an association function, a mapping from places to real values in [0,1]. $\beta: P\rightarrow D$ is an association function, a bijective mapping from laces to propositions.

 Let $A$ be a set of directed arcs. If $p_{j} \in I(t_{i})$, then there exists a directed arc $a_{ji} (\in A)$ from the place $p_{j}$ to the transition $t_{j}$. if $p_{k} \in O(t_{i})$, then there exists a directed arc $a_{ik} (\in A)$ from the transition $t_{i}$ to place $p_{k}$. If $f(t_{i})=\mu_{i}$, $(\mu_{i} \in [0,1])$, then the transition $t_{i}$ is said to be associated with a real value $\mu_{i}$. If $\beta(p_{i})=d_{i}, (d_{i} \in D)$, then the place $p_{i}$ is said to be associated with the proposition $d_{j}$.

The token value in a place $p_{i} (\in P)$ is denoted by $\alpha(p_{i})$, where $\alpha(p_{i}) \in [0.1]$. If $\alpha(p_{i}) = y_{i} (y_{i} \in [0,1])$ and $\beta(p_{i})=d(_{i})$, then it indicates that the degree of truth of proposition $d_{i}$ is $y_{i}$.

\begin{figure}[!tb]
\begin{center}
\includegraphics[scale=0.6]{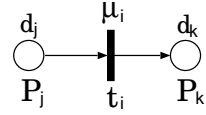}
\caption{Fuzzy Petri Net of If-Then rule}
\vspace{-0.5cm}
\label{fig:FPN_simple}
\end{center}
\end{figure}

By using a FPN, the fuzzy production rule such as Eq.(\ref{eq:Rule-0}) can be modeled as shown in Fig.\ref{fig:FPN_simple}. The FPN model has places, transitions, and tokens. A transition may be enabled to fire. A transition $t_{i}$ is enabled if for all $p_{j} (\in I(t_{i})), \alpha(p_{j})\geq \lambda$, where $\lambda$ is a threshold value in $[0,1]$. A transition $t_{i}$ fires by removing the tokens from its input places and then depositing one token into each of its output places. The token value in an output place of $t_{i}$ is calculated by using Eq.(\ref{eq:token}).

\begin{equation}
y_{k}=y_{i} \cdot \mu_{i}
\label{eq:token}
\end{equation}
If there are 2 or more fuzzy variables in the antecedent part of rules, the production of $\min$ of them and the transition is interpreted by the fuzzy reasoning as follows.
\begin{equation}
y_{k}=\min (y_{j1}, \cdots, y_{jn}) \cdot \mu_{ij}
\label{eq:Min}
\end{equation}

\begin{figure}[tbp]
\begin{center}
\subfigure[Type1]{
\includegraphics[scale=0.6]{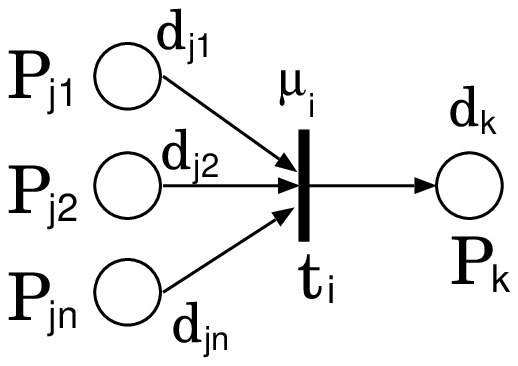}
\label{fig:FPN_1}
}
\subfigure[Type2]{
\includegraphics[scale=0.6]{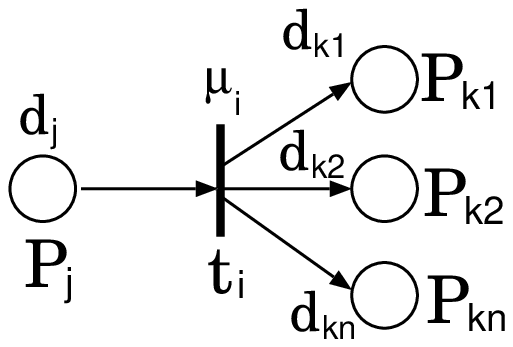}
\label{fig:FPN_2}
}
\subfigure[Type3]{
\includegraphics[scale=0.6]{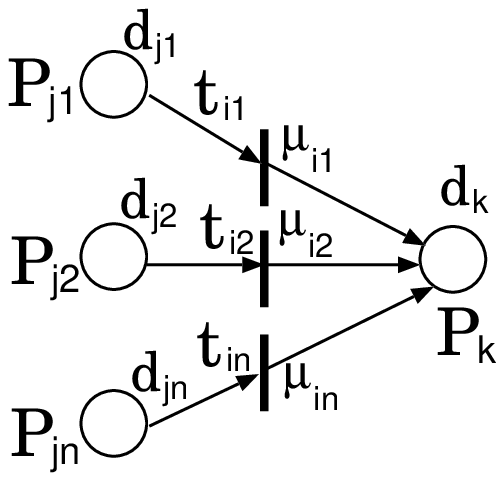}
\label{fig:FPN_3}
}
\subfigure[Type4]{
\includegraphics[scale=0.6]{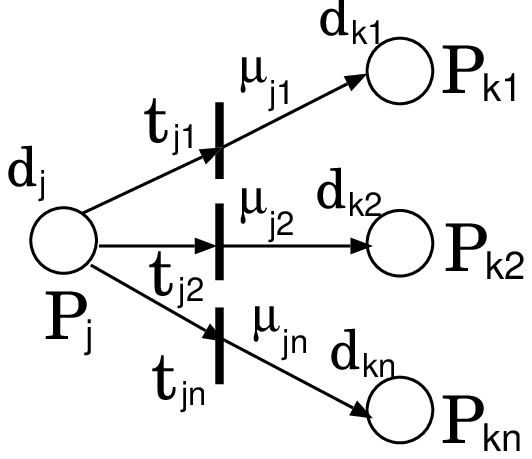}
\label{fig:FPN_4}
}
\caption{FPN models}
\label{fig:FPN_models}
\end{center}
\end{figure}

Furthermore, the following 4 types of IF-THEN rules by extending Eq.(\ref{eq:Rule-0}) are defined in Eq.(\ref{eq:FPN_rules}). Fig.\ref{fig:FPN_models} shows the FPN model of Eq.(\ref{eq:FPN_rules}). However, Type4 cannot derive clear implication, and then we don’t consider in this paper. By Eq.(\ref{eq:Min2}), the token values are calculated respectively.

\begin{eqnarray}
\nonumber TYPE 1:&&\; {\rm IF}\; d_{j1}\; {\rm and}\; d_{j2}\; \cdots\; {\rm and}\; d_{jn}\; {\rm Then}\; d_{k}, CF=\mu_{i}\\
\nonumber TYPE 2:&&\; {\rm IF}\; d_{j}\; {\rm Then}\; d_{k1}\; {\rm and}\; d_{k2}\; \cdots\; {\rm and}\; d_{kn},\! CF = \mu_{i} \\
\nonumber TYPE 3:&&\; {\rm IF}\; d_{j1}\; {\rm or}\; d_{j2}\; \cdots\; {\rm or}\; d_{jn}\; {\rm Then}\; d_{k},\\
\nonumber \; &&\!CF = \{\mu_{i1}, \mu_{i2}, \cdots, \mu_{in}\} \\
\nonumber TYPE 4:&&\; {\rm IF}\; d_{j}\;  {\rm Then}\; d_{k1}\; \cdots\; {\rm or}\; d_{k2}\; \cdots\; {\rm or}\; d_{kn},\\
\; &&\!CF = \{\mu_{j1}, \mu_{j2}, \cdots, \mu_{jn}\}
\label{eq:FPN_rules}
\end{eqnarray}
\begin{eqnarray}
\nonumber y_{k} &=& \min( y_{j1}, y_{j2}, \cdots, y_{jn} ) \cdot \mu_{i}\\
\nonumber y_{kl} &=& y_{j} \cdot \mu_{i}, \: (1 \leq l \leq n )\\
y_{k} &=& \max( y_{j1} \cdot \mu_{i1}, y_{j2} \cdot \mu_{i2}, \cdots, y_{jn} \cdot \mu_{in})
\label{eq:Min2}
\end{eqnarray}

\section{Hiroshima Tourist Information}
\label{sec:HiroshimaTouristInformation}
\subsection{Reasoning Rules}
\label{sec:HiroshimaTouristInformation_rules}
We have developed Android EGC application software which the agent works to evaluate the user's feelings in the conversation. Our developed application is tourist information system which estimates the user's feelings by EGC. The system can recommend the sightseeing spot, the local food corresponded to the user's feeling. For example, the system can guide the spot where the user feels ``happy'', even if he/she feels ``disgust.'' Our developed system for Hiroshima Tourist can guide some spots, local food shops, and local gifts. The system decides the next recommendation related to spots and foods by the reasoning of Fuzzy Petri Net to be more smooth communication between human and smartphone. We developed Fuzzy production rules to serve the hospitality in the concierge system.

The general tourist information system provides the unimaginatively conventional data due to the user's requirements. Our developed system tries to find out what the user wants to do under the consideration on his/her emotion. The morphological analyzing \& parsing process as shown in Fig. \ref{fig:processgeneratingemotion} is executed before the implementation of EGC. If the specified verb exists in in the sentence, the system reacts through the conversation. The specified verbs based on the user's behavior are $\{$`see,' `go,' `come,' `eat,' `(be) hungry,' `buy,' `look up'$\}$ and their synonyms. If the object of user's requirement is a sightseeing spot, that is, if the sentence includes verbs which mean behaviors related to place such as `see,' `go,' and `come,' the system measures the difference of 20 elements between the user's current emotion by EGC and the impression from the place. The impression was the normalized average on the results of questionnaire which asks the impression value for each spot by five-point scale. The impression value at the place means the degree and the kinds of emotion that the people images the place. In this paper, the impression values for 10 sightseeing spots in Hiroshima are investigated. If the current emotion type is classified into the negative emotion group, then the system recommends the spot of to being opposite values in negative emotion group. Otherwise, it recommends the spot of smallest difference. Moreover, if the object of user's requirement behaviors is not sightseeing spot, and if it is the other behaviors such as 'hungry,' `buy,' `look up,' and `see (not place),' the system detects the noun, and then the emotion value is calculated by using $FV$ in EGC. To summarize such ideas, the rules generated from EGC are classified into 3 categories related to the verbs and are defined as follows.

\begin{enumerate}
\item[\textbullet] Case 1) The sentence includes the verb: `Go', `Come', `See', or `Look for'.
\end{enumerate}
\begin{enumerate}[Rule 1)]
\item {\rm If} ``$Obj$ is $Spot$'' {\rm and} ``$AV$ of $Emotion$ is high'' {\rm Then} ``Recommend $Spot$''. {\rm If} ``$Obj$ is $Spot$'' {\rm and} ``$AV$ of $Emotion$ is not high'' {\rm Then} ``Recommend another $Spot$''.
\item {\rm If} ``$Obj$ is $Spot$'' {\rm and} ``$Emotion$ is negative,'' {\rm Then} ``Recommend the other $Spot$s where emotion becomes positive''.
\item {\rm If} ``$Obj$ is $Food$ {\rm or} $Gift$'' {\rm and} ``$FV$ is positive,'' {\rm Then} ``Recommend $Food$ {\rm or} $Gift$''.
\item {\rm If} ``$Obj$ is Nothing'' {\rm and} ``$AV$ of $Emotion$ is high,'' {\rm Then} ``Select a few $Spot$s with small difference''.
\item {\rm If} ``$Obj$ is Nothing'' {\rm and} ``$Emotion$ is negative,'' {\rm Then} Go to case3).
\begin{enumerate}
\setlength{\leftskip}{-1cm}
\item[\textbullet] Case 2) The sentence includes the verb: `Eat', or `Buy'.
\end{enumerate}
\item {\rm If} ``$FV$ is positive,'' {\rm Then} ``$Recommend$ $Food$ {\rm or} $Gift$''.
\begin{enumerate}
\setlength{\leftskip}{-1cm}
\item[\textbullet] Case 3) The sentence includes the verb: `talk'.
\end{enumerate}
\item {\rm If} ``$Emotion$ is negative,'' {\rm Then} Continue 'talk' till the dislike word is found. The words is recorded in the taboo list. Te system guides to talk another spots or foods and so on without the words in the taboo list.
\item {\rm Otherwise}, Go to Case 1) or Case 2),
\end{enumerate}
where $Obj$ is the object in case-frame representation, $Spot$ is the sightseeing spot, $Food$ and $Gift$ are the specialty of Hiroshima. $Emotion$ is the calculation result by using EGC, $FV$ means the favorite value in EGC. $AV$ means the agreement value between 20 kinds of user's emotion value and the impression value at each sightseeing spot. $Recommend$ means not only the proposal of the concrete things without words in the taboo list but also the consideration of its strength of recommendation.

\begin{figure}[tbp]
\begin{center}
\subfigure[Emotion Value]{
\includegraphics[scale=0.5]{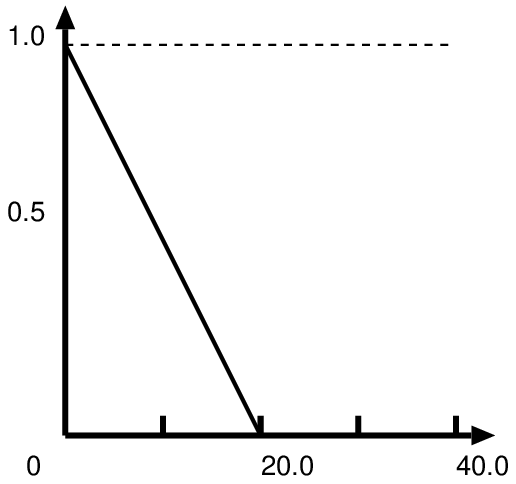}
\label{fig:Membership_EGC}
}
\subfigure[Favorite Value]{
\includegraphics[scale=0.5]{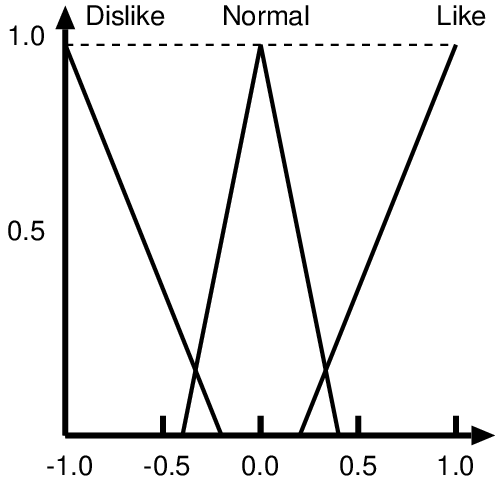}
\label{fig:Membership_FV}
}
\subfigure[Recommendation]{
\includegraphics[scale=0.5]{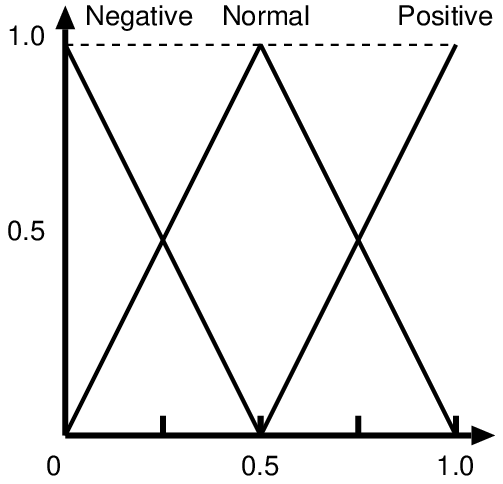}
\label{fig:Membership_Recommend}
}
\caption{Membership Function}
\label{fig:Membership}
\end{center}
\end{figure}

Fig.\ref{fig:Membership} shows the membership function of $AV$ of $Emotion$, $FV$, $Recommend$, respectively. Fig.\ref{fig:Membership_EGC} shows that $AV$ becomes smaller if $Emotion$ in x-axis is larger. In Fig.\ref{fig:Membership_FV}, there are 3 kinds membership functions, `Dislike,' `Normal,' and `Like.' In Fig.\ref{fig:Membership_Recommend}, the consequent part in the production rules are formed `Negative Recommend,' `Normal Recommend,' and `Positive Recommend.' The production rules for Hiroshima concierge are formed by the above mentioned membership function, so that the system can make a smooth conversation for the users.

\begin{figure}[tb]
\begin{center}
\includegraphics[scale=0.6]{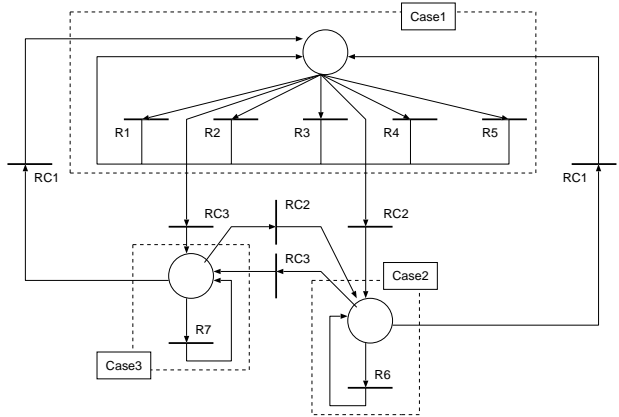}
\caption{Reasoning Process by Fuzzy Petri Net model}
\label{fig:FPNReasoning}
\end{center}
\end{figure}

Fig.\ref{fig:FPNReasoning} shows the reasoning process by Fuzzy Petri Net model based on the production rules for Hiroshima concierge. The model is divided into 3 parts corresponding to each case in the production rules. The connection denoted as `RC1', `RC2', and `RC3' plays the transition among 3 cases. The symbol such as `R1' means the number of production rules. In the model, the reasoning for sightseeing spots, the recommendation of local foods or gifts, and the consideration of mental condition are implemented in parallel distributed processing. 

\subsection{Experimental Results}
\label{sec:HiroshimaTouristInformation_results}
\begin{figure}[tbp]
\begin{center}
\subfigure[Miyajima]{
\includegraphics[scale=0.45]{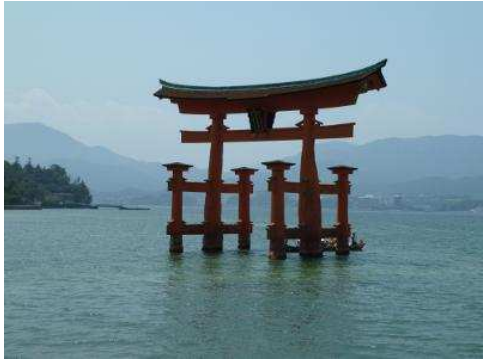}
\label{fig:Place_Miyajima}
}
\subfigure[Atomic Bomb Dome]{
\includegraphics[scale=0.45]{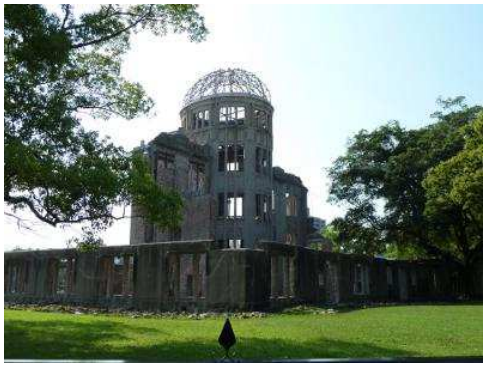}
\label{fig:Place_AtomicBomb}
}
\subfigure[Hiroshima Castle]{
\includegraphics[scale=0.45]{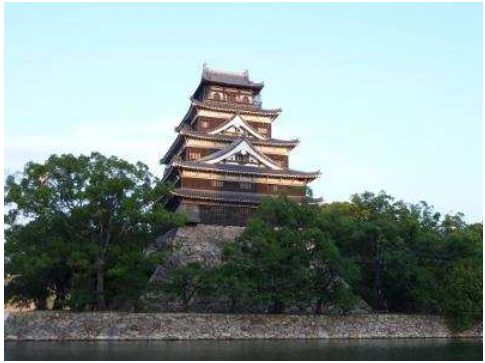}
\label{fig:Place_Castle}
}
\subfigure[Shukukeien]{
\includegraphics[scale=0.45]{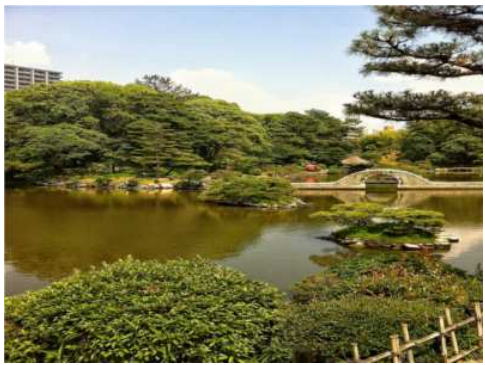}
\label{fig:Place_Shukukeien}
}
\caption{Hiroshima Sightseeing spots}
\label{fig:SightseeingSpots}
\end{center}
\end{figure}

Fig.\ref{fig:SightseeingSpots} shows some pictures of sightseeing spots which guide in Hiroshima Tourist Information system. In this paper, we prepared 10 places in Hiroshima and its environments, because the representative value of mental image is required to investigate by using questionnaire method.

\begin{figure}[tb]
\begin{center}
\includegraphics[scale=0.25]{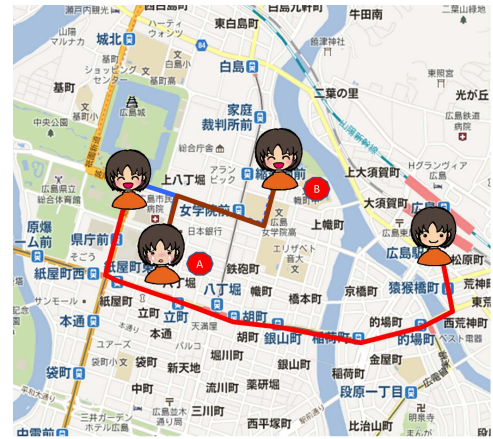}
\caption{Hiroshima Tourist Information Concierge System}
\label{fig:concierge}
\end{center}
\end{figure}

Fig.\ref{fig:concierge} shows the simulation result to act as guide from Hiroshima station to the downtown. The user starts from Hiroshima station in the right side in Fig.\ref{fig:concierge} and goes to the downtown in the center in Fig.\ref{fig:concierge}. Then the user reaches the first destination, Hiroshima Castle
 as shown in Fig.\ref{fig:Place_Castle}. Next, after the user talks with the system, the concierge recommends the restaurant in $\textcircled{\footnotesize A}$ of Fig.\ref{fig:concierge}. However, the user met temporary closure of the restaurant. The user feels sad and talks to the concierge system. The system can recognize the user's mental state and then recommend the favorite local food, `Okonomi-Yaki' and the restaurant near the spot $\textcircled{\footnotesize A}$. The recommended restaurant is near Shukukeien as shown in the symbol $\textcircled{\footnotesize B}$ in Fig.\ref{fig:concierge}. Shukukeien is a traditional Japanese landscape garden. The user can have a good meal before the sightseeing, and then can enjoy a comfortable sightseeing by the end of the day.

\section{Conclusion}
\label{sec:ConclusiveDiscussion}
The smartphone can use various kinds of application such as web browser, e-mail, Google map and so on. Especially, the voice recognition function is the outstanding application to spread the capability of mobile phone, because the current dialog system requires the user's typing. For example, the concierge system uses voice recognition function. However, the essential quality of dialog is limited to question and answer, although the recognition rate becomes good. In order to enjoy real conversation, the system can evaluate the user's emotion by using Android EGC\cite{Ichimura12}. Furthermore, we also developed the concierge system for tourists to implement the fuzzy Petri net model.

 Many tourists seem to come all the way from Kyoto, Osaka, and Tokyo to Hiroshima. However, most of them feels very regretful because of the few sightseeing spots in Hiroshima. In order to improve such a problem, Hiroshima Prefecture and the tourist association developed the smartphone application called `Hiroshima Quest' which enables travelers to plan enjoyable trip by bookmarking information of tourist sites, writing review and communicating with other users. Moreover, the Hiroshima tourist website replenishes the variety of information for foreigners. The Android application `Hiroshima Tourist map'\cite{Android_Market} has been developed to one of Mobile Phone based Participatory Sensing system, because not only tourism association but the local citizens should give the innovative and attractive information in sightseeing to visitors. We will embed the concierge system into our developed `Hiroshima Sightseeing map' in near future. The usability for the developed system will be investigated to put the system to practical use.

\section*{Acknowledgment}
This work was supported by JSPS KAKENHI Grant Number 25330366.

\end{document}